\documentclass[11pt,twoside]{article}

\usepackage{asp2006}
\usepackage{fancyhdr,graphicx,caption,rotating,multirow,lscape}
\DeclareGraphicsExtensions{.eps,.eps.gz,.ps,.jpg,.tiff}

\begin{document}

\title{Transiting Planet Simulations from the All Sky Extrasolar Planets Survey}

\author{Stephen R. Kane \& Jian Ge}
\affil{Department of Astronomy, University of Florida, 211 Bryant Space Science
Center, Gainesville, FL 32611-2055, USA}

\begin{abstract}
Many of the planets discovered via the radial velocity technique are hot
Jupiters in 3--5 day orbits with $\sim ~10$\% chance of transiting their
parent star. However, radial velocity surveys for extra-solar planets
generally require substantial amounts of large telescope time in order to
monitor a sufficient number of stars due to the single-object capabilities
of the spectrograph. A multi-object Doppler survey instrument has been
developed which is based on the dispersed fixed-delay interferometer design.
We present simulations of the expected results from the Sloan Doppler survey
based on calculated noise models and sensitivity for the instrument and
the known distribution of exoplanetary system parameters. We have developed
code for automatically sifting and fitting the planet candidates produced
by the survey to allow for fast follow-up observations to be conducted. A
transit ephemeris is automatically calculated by the code for each candidate
and updated when new data becomes available. The techniques presented here
may be applied to a wide range of multi-object planet surveys.
\end{abstract}

\section{Introduction}

Of the transiting planets known thus far, three were discovered via the radial
velocity technique and subsequently followed up photometrically. These are HD
209458b \citep{cha00}, HD 149026b \citep{sat05}, and HD 189733b \citep{bou05}.
A multi-object Doppler survey instrument, the W.M. Keck Exoplanet Tracker, has
been developed which builds upon the success \citep{ge06} of the prototype
instrument, Exoplanet Tracker (ET), which is based on the dispersed fixed-delay
interferometer design. This new instrument is being used to conduct an All Sky
Extrasolar Planets Survey (ASEPS) with the Sloan 2.5m wide-field telescope,
expected to dramatically increase the detection rate using the Doppler method.
Since the initial survey is optimised towards the discovery of hot Jupiters
orbiting stars in the magnitude range $8 < V < 12$, the planet discoveries will
be ideal candidates for the possible detection of a transit signature in the
corresponding lightcurve. We have conducted survey simulations \citep{kan06b}
to estimate how many transiting planets can be expected.

\section{Simulated Data}

To estimate how many planets we expect to detect in a given radial velocity
observing program, we performed a series of Monte-Carlo simulations which inject
planets into a realistic sample of target stars based on the known distribution
and characteristics of exoplanets. For this simulation, a stellar population
model was generated using the Besancon Galactic model \citep{rob03} for a
magnitude limited survey in the Kepler field. The Kepler field is an ideal
location for a multi-object radial velocity survey since the results would not
only compliment the transit survey to be undertaken by the Kepler mission, but
also aid in target selection. In total, 23708 main sequence stars were used in
the simulation from which 751 were concluded to harbour planets based on the
planet-metallicity correlation \citep{fis05}. The cumulative histogram of the
planet-harbouring probability for the sample, shown in Figure 1 (left),
demonstrates the substantial reduction in probability beyond $\sim 5$\%.

\begin{figure*}
  \begin{center}
    \begin{tabular}{cc}
      \includegraphics[angle=270,width=6.0cm]{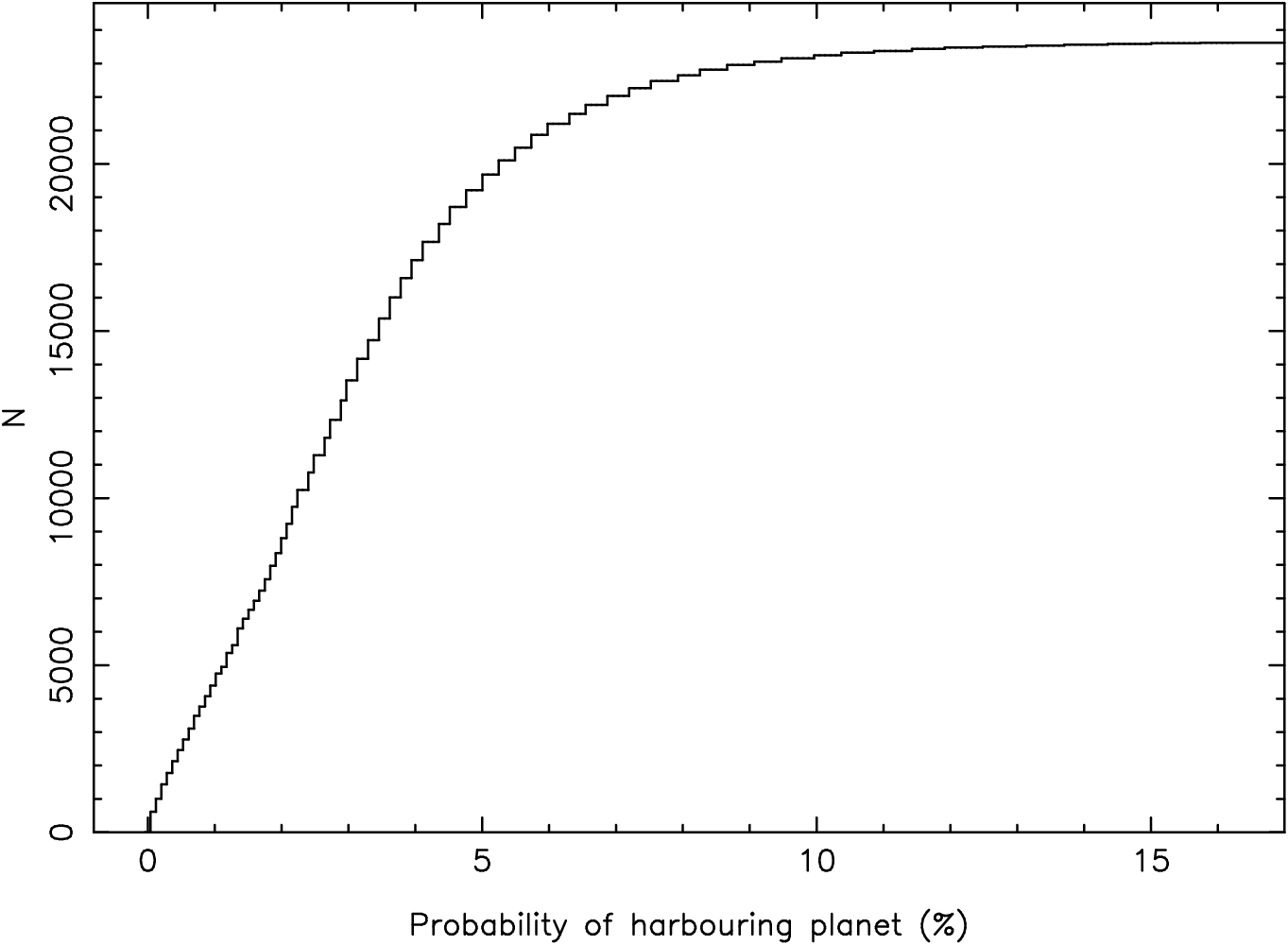} &
      \includegraphics[angle=270,width=6.0cm]{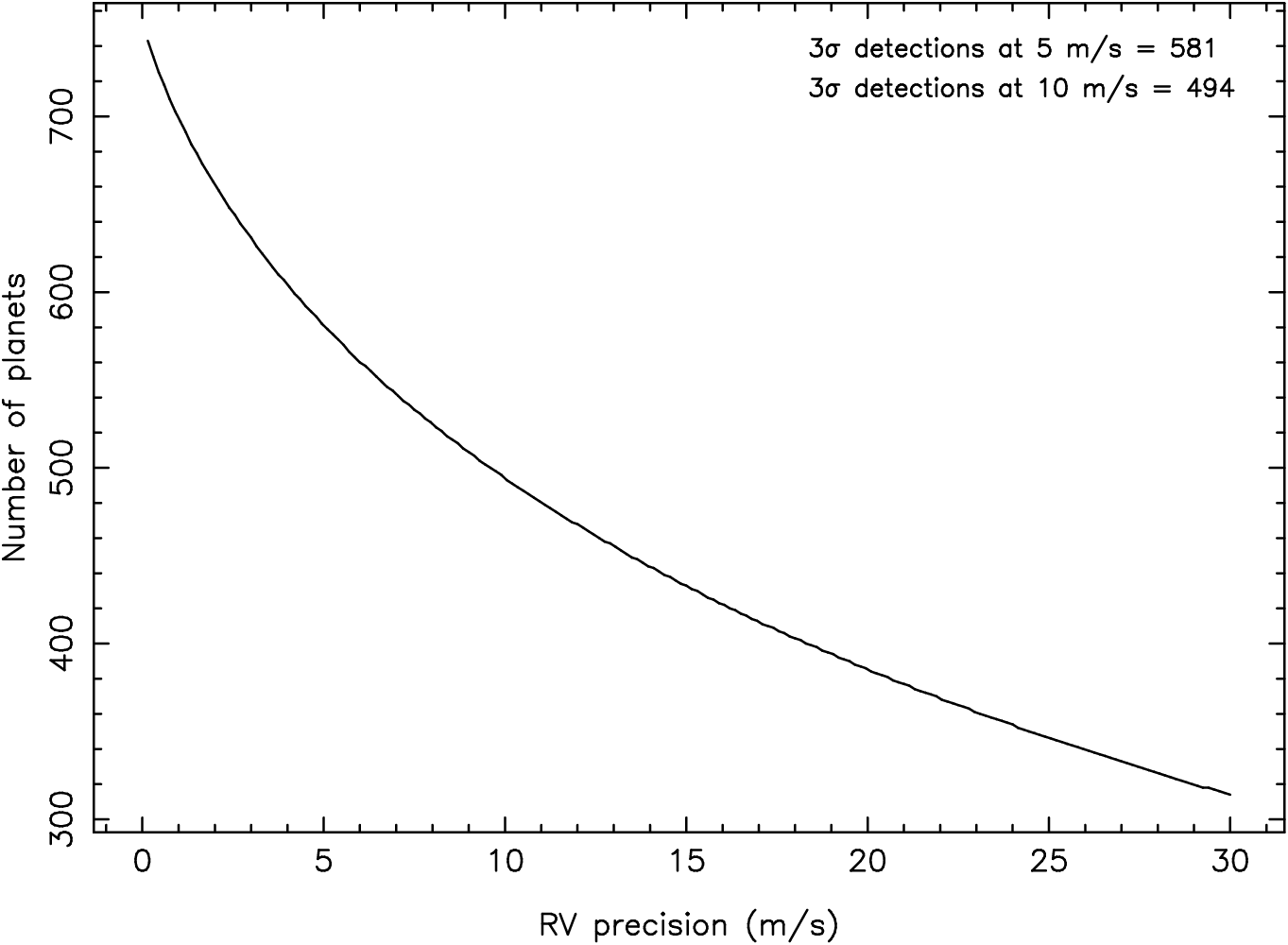} \\
     \end{tabular}
  \end{center}
  \caption{Cumulative planet-harbouring probability histogram (left) and the
    number of detectable planets as a function of RV precision (right).}
\end{figure*}

The distribution of planet parameters was used to calculate the expected
distribution of radial velocity amplitudes. Figure 1 (right) shows the number
of planets detectable from the simulated data at the 3$\sigma$ level as a
function of the rms radial velocity precision of the experiment. In this
example, doubling the precision of the instrument increases the planet yield by
$\sim 17$\%, whereas doubling the number of survey stars will increase the
planet yield by 100\%.

\section{Radial Velocity Fitting Code}

The FORTRAN code written for the purpose of sifting planet candidates from the
dataset uses a weighted Lomb-Scargle (L-S) periodogram to detect a periodic
signal in the data. If a significant periodic signal is detected, the data are
subjected to an iterated grid-search fitting routine to determine the best fit
planetary parameters for that target. Figure 2 (left) shows a typical
periodogram where the dotted lines indicate various false-alarm probabilities.

\begin{figure*}
  \begin{center}
    \begin{tabular}{cc}
      \includegraphics[angle=270,width=6.0cm]{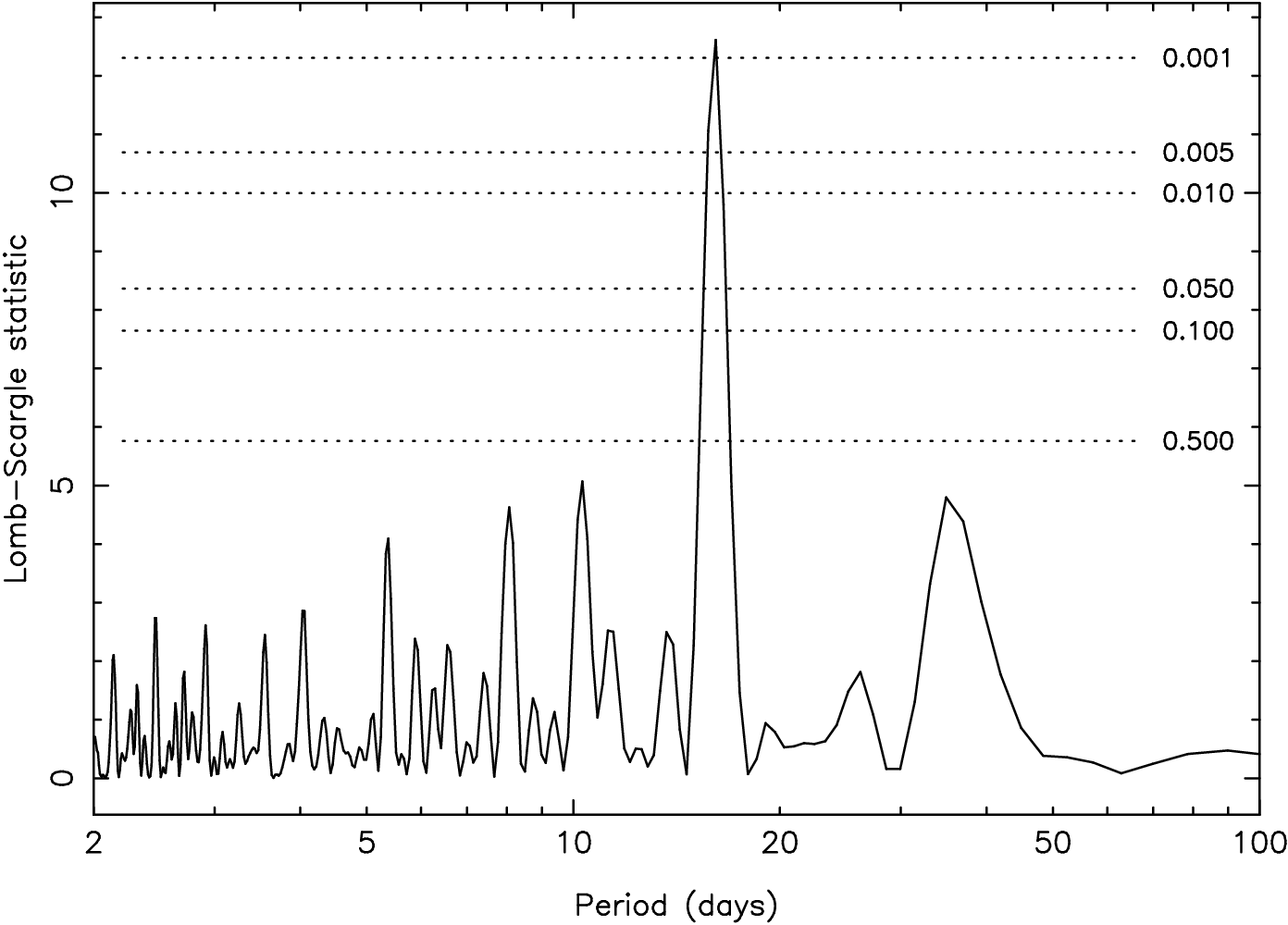} &
      \includegraphics[angle=270,width=6.0cm]{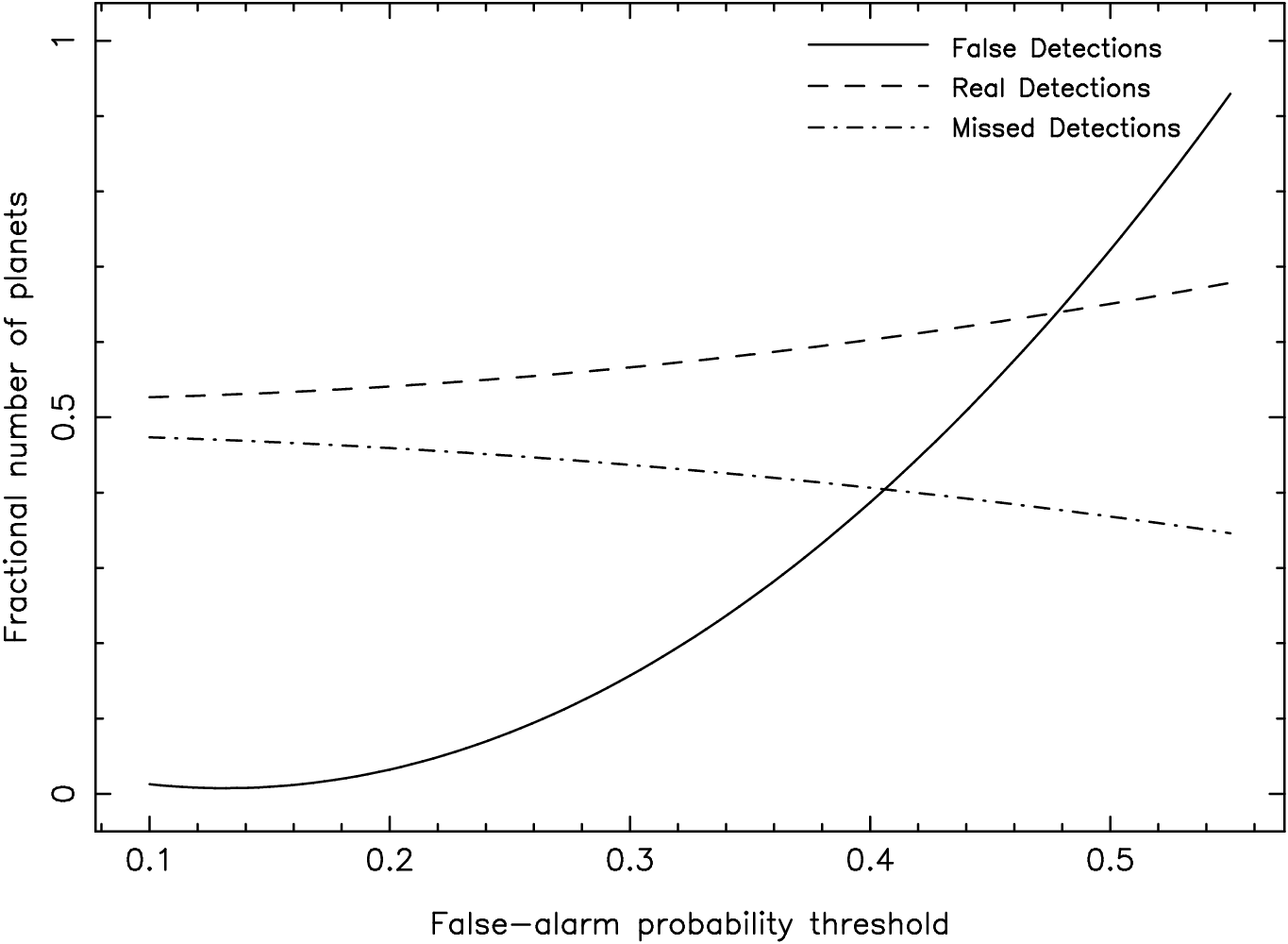} \\
     \end{tabular}
  \end{center}
  \caption{Periodogram for data containing a planetary signature (left)
    and the planets recovered as a function of the false-alarm probability
    threshold (right).}
\end{figure*}

The number of false detections resulting from this technique depends upon the
periodic false-alarm probability threshold one adopts as the selection criteria.
Figure 2 (right) shows that by adopting a relatively conservative threshold,
one can eliminate false detections with minimal loss to planet yield.

Those planetary signatures failing to be recovered are generally at the photon
noise-limit of the instrument or have a period considerably more than the survey
duration. This sifting and fitting method has proven to be remarkably robust and
is able to operate in an automated fashion.

\section{Number of Transiting Planets and Photometric Follow-up}

The total planet yield expected from the simulated data for the Kepler field
survey is $\sim 275$ planets for an instrument such as the ASEPS instrument
assuming a 30 day observing window. To approximate the number of transiting
planets, the geometric transit probability was calculated for each star/planet
system as part of the Monte-Carlo simulation. Once all the detection limitations
are taken into account, the total expected number is $\sim 10$ transiting
planets.

Considering the high number of expected hot Jupiters expected from the survey,
the radial velocity fitting code automatically calculates a transit ephemeris for
each candidate and is updated when new data becomes available. Since the transit
duration is brief compared with the fitted period, the maximum window for
obtaining photometric transit observations after the radial velocity data has
been obtained is also calculated. Through this process, we hope to provide a quick
and efficient method for maximising the number of transit discoveries
\citep{kan06a}.

\acknowledgements
We acknowledge support from the W.M. Keck Foundation, NSF, NASA, and the
University of Florida.

\end{document}